\documentclass[aps,twocolumn]{revtex4}
\usepackage{amsmath}
\usepackage{graphicx}
\usepackage{floatflt,epsfig}
\usepackage{lscape}
\usepackage{ulem}
\usepackage{natbib}
\usepackage{hyperref}
\begin{document}

\title{Preliminary EoS for core-collapse supernova simulations
     with the QMC model}
\author{Guilherme Grams}
\affiliation{Universidade Federal de Santa Catarina, Brazil}
\author{Alexandre M. Santos}
\affiliation{Universidade Federal de Santa Catarina, Brazil}
\author{D\'ebora P. Menezes}
\affiliation{Universidade Federal de Santa Catarina, Brazil}
\begin{abstract}
In this work we present the preliminary results of a complete equation of state (EoS) 
for core-collapse supernova simulations. 
We treat uniform matter made of nucleons using the the quark-meson coupling 
(QMC) model. We show a table with a variety of thermodynamic quantities, which covers 
the proton fraction range $Y_{p}=0-0.65$ with the linear grid spacing $ \Delta Y_{p}=0.01$ ($66$
points) and the density range $\rho _{B}=10^{14}-10^{16}$g.cm$^{-3}$ with the logarithmic grid 
spacing $\Delta $log$ _{10}(\rho _{B}/[$g.cm$^{-3}])=0.1 $ ($21$
points). This preliminary study is performed at zero temperature and
our results are compared with the widely used EoS already available in
the literature.
\end{abstract}


\maketitle

\section{Introduction}

Although the theory related to supernova (SN)
has made a remarkable progress in the past decade, there are many questions
still to be answered. The catastrophic infall of the core of a massive star, reversed to trigger
the powerful ejection of the stellar mantle and envelope in a supernova explosion, was identified 
as the crucial role in the synthesis of heavy elements. But when, why and how it
happens is a fundamental problem of stellar astrophysics that remains to be explained.
The implosion of stellar cores was also proposed as part of the scenario of the stellar death \cite{revsn,hoyle}.

There are no doubts that supernovae explosions are an unique
phenomenon in nature and an excellent laboratory to test extreme 
physics conditions. Unfortunately SN tends to happen once in a century per galaxy
which makes the study of these \textit{great labs} very difficult. Due to this observational difficulty,
simulations of core-collapse supernova have played an important role
in the study of supernovae explosions and their possible remnants.
The equation of state of nuclear physics is a fundamental ingredient in the simulation of SN explosion.
Simulations of core-collapse supernovae have to be fed with a wide
range of thermodynamic conditions. Moreover, extremely
high density and temperature may be achieved when black holes are formed by failed supernovae. 
To date, the temperature is believed to vary from zero to more than $100$ MeV,  densities from $10^{5}$ to more than $10^{15}$ g.cm$^{-3}$
and the proton fraction up to about $0.6$. Fulfilling these
conditions makes the construction of a complete EoS a very hard work,
mainly at low densities, where a variety of sub-structures and light
clusterization are possible. For these reasons, there are only few
complete EoS available in the literature. The most commonly used EoS
are those of 
Lattimer and Swesty \cite{lattimer}, Shen \cite{shen98, shen11}
and Hempel \cite{hempel}.

The Lattimer \cite{lattimer} EoS is based on a
compressible liquid drop model with a Skyrme force for nucleon interactions. The Shen EoS from 1998 \cite{shen98} 
was the first equation of state for supernova simulations using a relativistic nuclear model. The upgrade to Shen's 
work was published in 2011 \cite{shen11} with more points in the table and the inclusion of the $\Lambda $ hyperons.
Both works developed by Shen were constructed with the relativistic
mean field Walecka model \cite{walecka} using the TM1 \cite{tm1} parameterization. 
Hempel's EoS \cite{hempel} is based on the TM1, TMA \cite{tma} and FSUgold \cite{fsugold} parameterizations of the
Walecka model and use the nuclear statistical equilibrium model of Hempel and Schaffner-Bielich \cite{hs}, which includes excluded volume effects. 

\begin{table*}[t]
  \centering

  \begin{tabular}{lcclllllllcccccccccc}
\hline
Model  \;\; &    M   \;\; & $m_q$ \;\; & \; $m_\sigma$\; &\; $m_\omega$\; & $m_\rho$ \;\; &  $g_\rho$ \;\; & $g_\omega$ \;\; &  $g_\sigma $ \;\; & $B_{N}^{1/4}$\;\;  & NLT 	 \;   &  DDP      \\

       \;\; &  (MeV)  \;\; & (MeV)\;\; & (MeV)     \;\; & (MeV)      \;\; & (MeV)    \;\; &           \;\; &            \;\; &              \;\; & (MeV)     \;\;  &     	 \;   &          \\
   
\hline
QMC    \;\; & 939.0 \;\;  & 5.5  \;\;  &   550    \;\;   &    783    \;\; &    770  \;\;  &  8.6510   \;\; &  8.9817    \;\; &  5.9810$^*$    \;\; &  210.85    \;\;  &    no       \;  &   no     \\

\hline
NL3    \;\; & 939.0 \;\;  &  --  \;\;  &    508.194 \;\; &  782.501  \;\; &   763   \;\;  &  8.9480   \;\; &  12.868   \;\;  &   10.217  \;\;      &   --    \;\;     &     yes    \; &   no   \\
  
GM1   \;\;  & 939.0  \;\; &  --  \;\;  &   550    \;\;   &   783    \;\;  &   770   \;\;  &  8.1945  \;\;  &  10.608    \;\; &   9.5684  \;\;       &  --    \;\;      &    yes    \; &   no \\

TM1   \;\;  & 938.0 \;\;  &  --  \;\;  &  511.197 \;\;   &  783     \;\;  &   770   \;\;  &  4.6321  \;\;  & 12.613    \;\;  &  10.028   \;\;       &  --    \;\;     &    yes   \;   &   no     \\

FSUgold\;\; & 939.0 \;\;  &  --   \;\; & 491.5    \;\;   &  782.5   \;\;  &    763  \;\;  & 11.7673   \;\; & 14.301     \;\; & 10.592   \;\;        & --    \;\;     &     yes   \;   &     no    \\

TMA$^{\dagger}$ \;\;  & 938.9 \;\;  &  --  \;\;  & 519.151  \;\;   &  781.95   \;\; &    768.1 \;\; & 3.800    \;\;   & 12.842    \;\; & 10.055  \;\;   & --   \;\;     &    yes      \;    &   no \\

TW      \;\;  & 939.0  \;\; &  --  \;\;  &   550    \;\;   &   783    \;\;  &   763   \;\;  & 7.32196$^{**}$ \;\; & 13.2901$^{**}$ \;\; & 10.728$^{**}$ \;\; &  --    \;\;     &    no      \; & yes      \\

\hline
 \end{tabular} 
\caption{Parameters used in the QMC model and  different Walecka model
  parameterizations. In the first line we present the parameters used
  in the present work with the QMC model. The NL3, GM1 and TW
  parameterizations of the Walecka model are 
used here for a comparison of bulk matter properties.  TM1 is the
parameterization used in Shen's work. The FSUgold and TMA were used in
Hempel's work. 
$^* g^{q}_{\sigma}$ is the quark-meson coupling in the QMC model. 
$^{**}$ values taken at saturation for the TW
parameterization. TMA$^{\dagger}$: the coupling parameters $g_i$ of
the set TMA are chosen to be mass-number dependent such that $g_i=a_i+b_i/A^{0.4}$,
with $a_i$ and $b_i$ being constants \cite{tma}; for infinite matter
as in the stellar matter, one has an infinite nucleus, and then 
the limit A $\mapsto$ infinity is taken so that 
$g_i=a_i$. NLT= Nonlinear terms. DDP= Density dependent parameters.)}
  \label{tab:parameters}
\end{table*}

An interesting analysis of the different parameterizations of the relativistic-mean-field (RMF) models was made by Dutra et al \cite{dutra14}, where
the authors analyzed 263 different RMF models under several constrains
related to the symmetric nuclear matter (SNM) and pure neutron matter 
(PNM). 
The TM1 parameterization used by Shen failed under six of this
constraints, the TMA used by Hempel failed in five, and the FSUgold,
also used by Hempel
failed in one constraint. More details of this analysis in \cite{dutra14}.

The works of Hempel, Lattimer and Shen were successful and very useful in many calculations in the 
last decade \cite{yuichiro, nasu, abdika, fryer, thompson03, pejcha}, but there are some simulations of SN in which the supernova does not
explode \cite{buras}. It is believed that these failures are due to some problems with the nuclear EoS.

In this work we present our preliminary results for the construction of an EoS grid for 
core-collapse supernova simulations, with the quark-meson coupling (QMC) model \cite{guichon}.
 
In the QMC model, nuclear matter is described as a system of non-overlapping MIT bags \cite{mitbag} that
interact through the exchange of scalar and vector meson fields. Many applications and extensions of the model have been made in the
last years \cite{fleck,saito,saito95,guicho96,panda}. It is found that
the EoS for infinite nuclear matter at zero temperature derived from
the QMC model is softer than the one obtained with the Walecka
model. This might be a problem if one wants to describe very massive 
compact objects \cite{demorest,antoniadis}, but as far as SN simulations are concerned, it is worth testing it because 
apart from numerical accordance, one can interpret that 
starting from quark degrees of freedom is an advantage on the physical meaning underneath.
Moreover, the effective nucleon mass obtained with the
QMC model lies in the range 0.7-0.8 of the free nucleon mass, which agrees with the results derived from the non-relativistic analysis
of scattering of neutrons from lead nuclei \cite{johnson} and is
larger in comparison with the effective mass obtained with some of the
different parameterizations of the Walecka models. A low effective
mass at saturation can be a problem when hyperons are included in the
calculation, as discussed in the next section.

All the few works for the purpose of the SN simulations that are based
on relativistic models, use different parameterizations of the 
Walecka model, in which the nucleons interact between each other
through the exchange of mesons. We believe that
with the quarks degree of freedom present in the QMC model, a more
fundamental physics lacking in the already used models, can be tested
and possibly contribute for the SN simulations to explode.

Another possible use of this preliminary EoS obtained with the QMC model is
the study of the cooling of compact stars, which serve as an important window 
on the properties of super-dense matter and neutron star structure,
and is very sensitive 
to the nuclear equation of state \cite{page,negreiros,fortin,carvalho}.

This paper is structured as follows: in the second section we present a review of the quark-meson 
coupling model. In the third section we present our results for the EoS. The last section is reserved for the final remarks and conclusions.

\section{The quark-meson coupling model}

In the QMC model, the nucleon in nuclear medium is assumed to be a
static spherical MIT bag in which quarks interact with the scalar ($\sigma$)
and vector ($\omega$, $\rho$) fields, and those
are treated as classical fields in the mean field
approximation (MFA) \cite{guichon}.
The quark field, $\psi_{q_{N}}$, inside the bag then
satisfies the equation of motion:
\begin{eqnarray}
\left[i\,\rlap{/}\partial \right.&-&(m_q^0-g_\sigma^q\,)-g_\omega^q\, \omega\,\gamma^0\nonumber \\
&+&\left. \frac{1}{2} g^q_\rho \tau_z \rho_{03}\gamma^0\right]
\,\psi_{q_{N}}(x)=0\ , \quad  q=u,d
\label{eq-motion}
\end{eqnarray}
where $m_q^0$ is the current quark mass, and $g_\sigma^q$, $g_\omega^q$ and $g_\rho^q$
denote the quark-meson coupling constants. The normalized ground state for a quark in the bag 
is given by
\begin{eqnarray}
\psi_{q_{N}}({\bf r}, t) &=& {\cal N}_{q_{N}} \exp 
\left(-i\epsilon_{q_{N}} t/R_N \right) \nonumber \\
&\times& \left(
\begin{array}{c}
  j_{0_{N}}\left(x_{q_{N}} r/R_N\right)\\
i\beta_{q_{N}} \vec{\sigma} \cdot \hat r j_{1_{N}}\left(x_{q_{N}} r/R_N\right)
\end{array}\right)
 \frac{\chi_q}{\sqrt{4\pi}} ~,
\end{eqnarray}
where
\begin{equation}
\epsilon_{q_{N}}=\Omega_{q_{N}}+R_N\left(g_\omega^q\, \omega+
\frac{1}{2} g^q_\rho \tau_z \rho_{03} \right) ,
\end{equation}
and,
\begin{equation}
\beta_{q_{N}}=\sqrt{\frac{\Omega_{q_{N}}-R_N\, m_q^*}{\Omega_{q_{N}}\, +R_N\, m_q^* }}\ ,
\end{equation}
with the normalization factor given by
\begin{equation}
{\cal N}_{q_{N}}^{-2} = 2R_N^3 j_0^2(x_q)\left[\Omega_q(\Omega_q-1)
+ R_N m_q^*/2 \right] \Big/ x_q^2 ~,
\end{equation}
where $\Omega_{q_{N}}\equiv \sqrt{x_{q_{N}}^2+(R_N\, m_q^*)^2}$,
$m_q^*=m_q^0-g_\sigma^q\, \sigma$, $R_N$ is the
bag radius of nucleon $N$ and $\chi_q$ is the quark spinor. The bag eigenvalue for nucleon $N$, $x_{q_{N}}$, is determined by the
boundary condition at the bag surface
\begin{equation}
j_{0_{N}}(x_{q_{N}})=\beta_{q_{N}}\, j_{1_{N}}(x_{q_{N}})\ .
\label{bun-con}
\end{equation}

The energy of a static bag describing nucleon $N$ consisting of three quarks in ground state
is expressed as
\begin{equation}
E^{\rm bag}_N=\sum_q n_q \, \frac{\Omega_{q_{N}}}{R_N}-\frac{Z_N}{R_N}
+\frac{4}{3}\,  \pi \, R_N^3\,  B_N\ ,
\label{ebag}
\end{equation}
where $Z_N$ is a parameter which accounts for zero-point motion
of nucleon $N$ and $B_N$ is the bag constant.
The set of parameters used in the present work is determined by enforcing 
stability of the nucleon (here, the ``bag''), much like in \cite{alex09}, 
so there is a single value for 
proton and neutron masses. The effective mass of a nucleon bag at rest
is taken to be $M_N^*=E_N^{\rm bag}.$

The equilibrium condition for the bag is obtained by
minimizing the effective mass, $M_N^*$ with respect to the bag radius
\begin{equation}
\frac{d\, M_N^*}{d\, R_N^*} = 0,\,\,\;\;\; N=p,n ,
\label{balance}
\end{equation}
By fixing the bag radius $R_N=0.6$ fm and the bare nucleon mass $M=939$ MeV the unknowns $Z_N=4.0050668$ and 
$B^{1/4}_N=210.85$MeV are then obtained.
Furthermore, the desired values of $ B/A \equiv   \epsilon/\rho - M = -15.7$~MeV at saturation
$n=n_0=0.15$~fm$^{-3}$, are achieved by setting $g_\sigma^q=5.9810$, $g_{\omega}=8.9817$,
where $g_\omega =3g^q_\omega$ and $g_\rho =3g^q_\rho$. All the parameters used in this work are 
shown in Table \ref{tab:parameters}.

The total energy density of the nuclear matter reads 
\begin{eqnarray}
\varepsilon = \frac{1}{2}m^{2}_{\sigma}\sigma+\frac{1}{2}m^{2}_{\omega}\omega^{2}_{0}+\frac{1}{2}m^{2}_{\rho}\rho^{2}_{03}\nonumber \\
+\sum_{N} \frac{1}{\pi^{2}}\int^{k_{N}}_{0}k^{2}dk[k^{2}-M^{*2}_{N}]^{1/2},
\label{energ dens}
\end{eqnarray}
and the pressure is,
%
\begin{eqnarray}
p = -\frac{1}{2}m^{2}_{\sigma}\sigma+\frac{1}{2}m^{2}_{\omega}\omega^{2}_{0}+\frac{1}{2}m^{2}_{\rho}\rho^{2}_{03}\nonumber \\
+\sum_{N} \frac{1}{\pi^{2}}\int^{k_{N}}_{0}k^{4}dk/[k^{2}-M^{*2}_{N}]^{1/2}.
\label{press}
\end{eqnarray}

The vector mean field $ \omega_0 $ and $ \rho_{03} $ are determined through
\begin{equation}
\omega_0 =\frac{g_\omega (n_p +n_n)}{m^{2}_{\omega}}, \; \rho_{03}=\frac{g_\rho (n_p -n_n)}{m^{2}_{\rho}},
\end{equation}
where
\begin{equation}
n_B=\sum_N \frac{2 k_{N}^3}{3 \pi ^2}, \quad N=p,n.
\end{equation}
is the baryon density. 

Finally, the mean field $\sigma$ is fixed by imposing that
\begin{equation}
\frac{\partial \varepsilon}{\partial \sigma}=0.
\end{equation}
%

\begin{figure}[!]
\begin{tabular}{l}
\includegraphics[width=9.cm,angle=0]{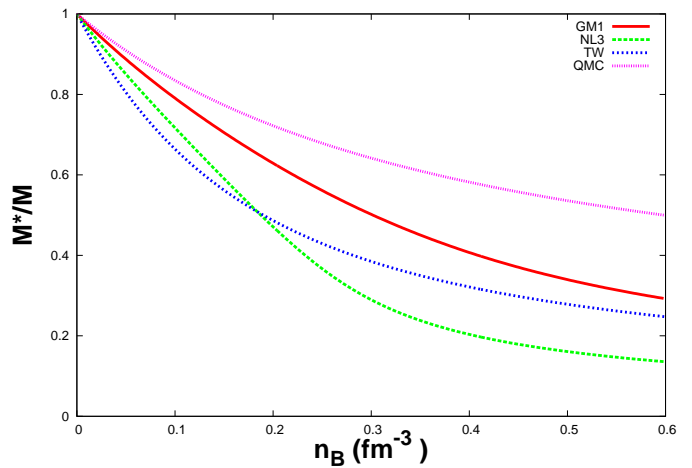}
\end{tabular}
\caption{The effective mass of the QMC model and three Walecka parameterizations.}
\label{fig1}
\end{figure}

\begin{figure}[!]
\begin{tabular}{l}
\includegraphics[width=9.cm,angle=0]{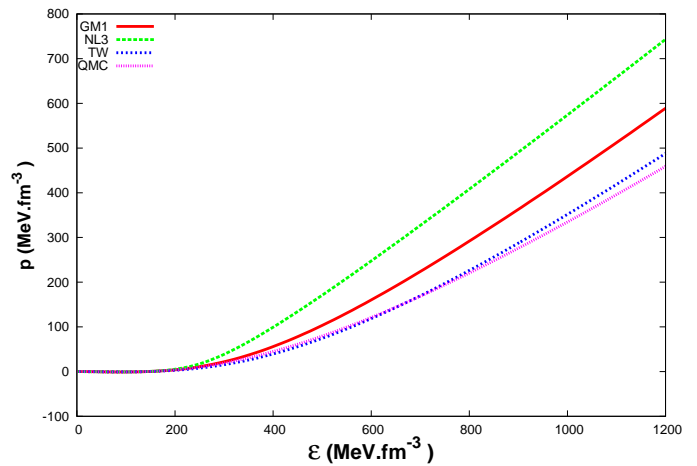}
\end{tabular}
\caption{Equation of state of the QMC model and three Walecka model parameterizations.}
\label{fig2}
\end{figure}

\begin{table}
\begin{tabular}{lclccc}
\hline
Model   &$B/A$ & $n_{0}$   & $M^*/M$  & $\mathcal E_{sym}$ &   K     \\
        & (MeV)&(fm$^{-3}$)&          & (MeV)              & (MeV)   \\
\hline
QMC    	&-15.7 &0.150      & 0.77     & 34.5               & 295     \\
\hline
NL3     &-16.2 &0.148      & 0.60     & 37.4               & 272  \\
GM1 	&-16.3 &0.153      & 0.70     & 32.5               & 300  \\
TM1     &-16.3 &0.145      & 0.63     & 36.8               & 281  \\
FSUgold	&-16.3 &0.148      & 0.62     & 32.6               & 230  \\
TMA     &-16.0 &0.147      & 0.63     & 30.7               & 318  \\
TW      &-16.2 &0.153      & 0.56     & 32.6               & 240  \\
\hline
\end{tabular}
\caption{Nuclear matter bulk properties obtained with the QMC model, two different parameterizations of Walecka model
and one density dependent model we use in this paper, and  the four parameterizations used in the works of Hempel
and Shen. All quantities are taken at saturation.} 
\label{tab:comp}

\end{table}



It is always important to check the behavior of the models in the
symmetric nuclear matter at saturation density and zero temperature, i.e., the bulk nuclear matter properties.
A comprehensive work in this direction is \cite{dutra14}, but the QMC model was not analyzed.
Therefore, we compare the QMC model with two parameterization of the well known
Walecka-type models, namely: GM1 \cite{gledenning} and NL3 \cite{nl3}
and the density dependent parameter model TW 
\cite{tw}.
In this work we have chosen GM1 for being a parameterization which
gives a good value for the effective mass, NL3 because it is a very
common 
standard parameterization and TW because it is a very good density
dependent parameterization according to \cite{dutra14}.
In \cite{dutra14}, it was found that GM1 failed under six constrains related to the symmetric nuclear matter (SNM) and pure
neutron matter (PNM), NL3 failed under nine and TW did not satisfy
only one constraint.

In Table \ref{tab:parameters} we can see the different parameters here analyzed and the ones used in the works of Shen \cite{shen98,shen11} 
and Hempel \cite{hempel}.  

\begin{figure}[!]
\begin{tabular}{l}
\includegraphics[width=9.cm,angle=0]{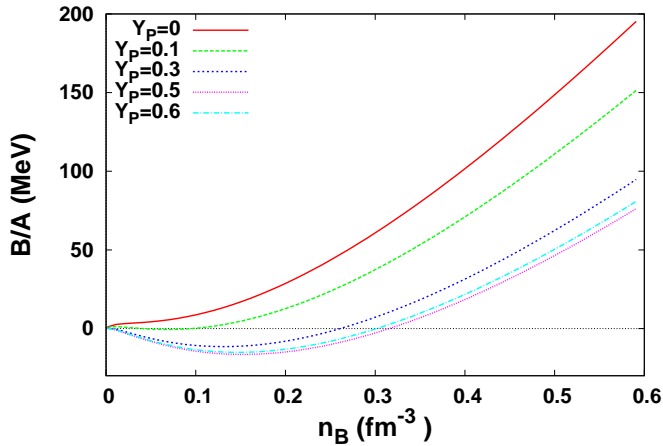}
\end{tabular}
\caption{Binding energy of the nucleons as function of the baryon density with the proton fractions
$Y_p=$0, 0.1, 0.3, 0.5 and 0.6.}
\label{fig3}
\end{figure}

\begin{figure}[!]
\begin{tabular}{l}
\includegraphics[width=9.cm,angle=0]{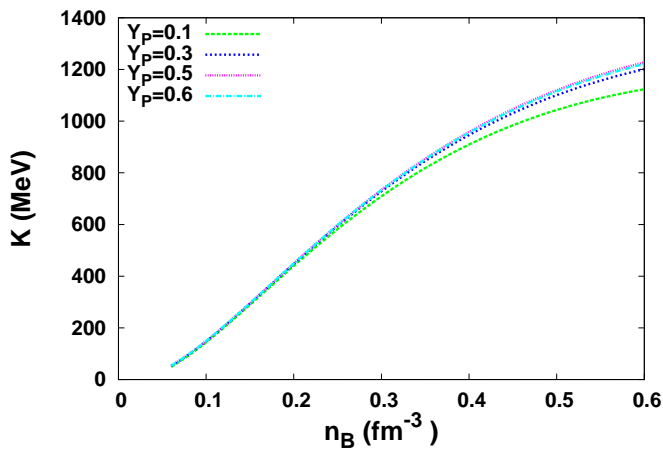}
\end{tabular}
\caption{Compression modulus as function of the baryon density with the proton fractions
$Y_p=$0.1, 0.3, 0.5 and 0.6.}
\label{fig4}
\end{figure}

\begin{figure}[!]
\begin{tabular}{l}
\includegraphics[width=9.cm,angle=0]{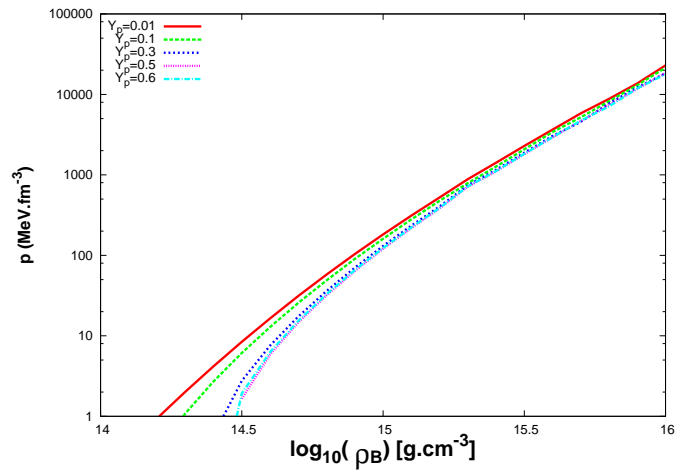}
\end{tabular}
\caption{The relation between pressure and $\rho_B$ for the $Y_p=$0.01, 0.1, 0.3, 0.5 and 0.6 protons fractions.}
\label{fig5}
\end{figure}

\begin{figure}[!]
\begin{tabular}{llll}
\includegraphics[width=8.cm,angle=0]{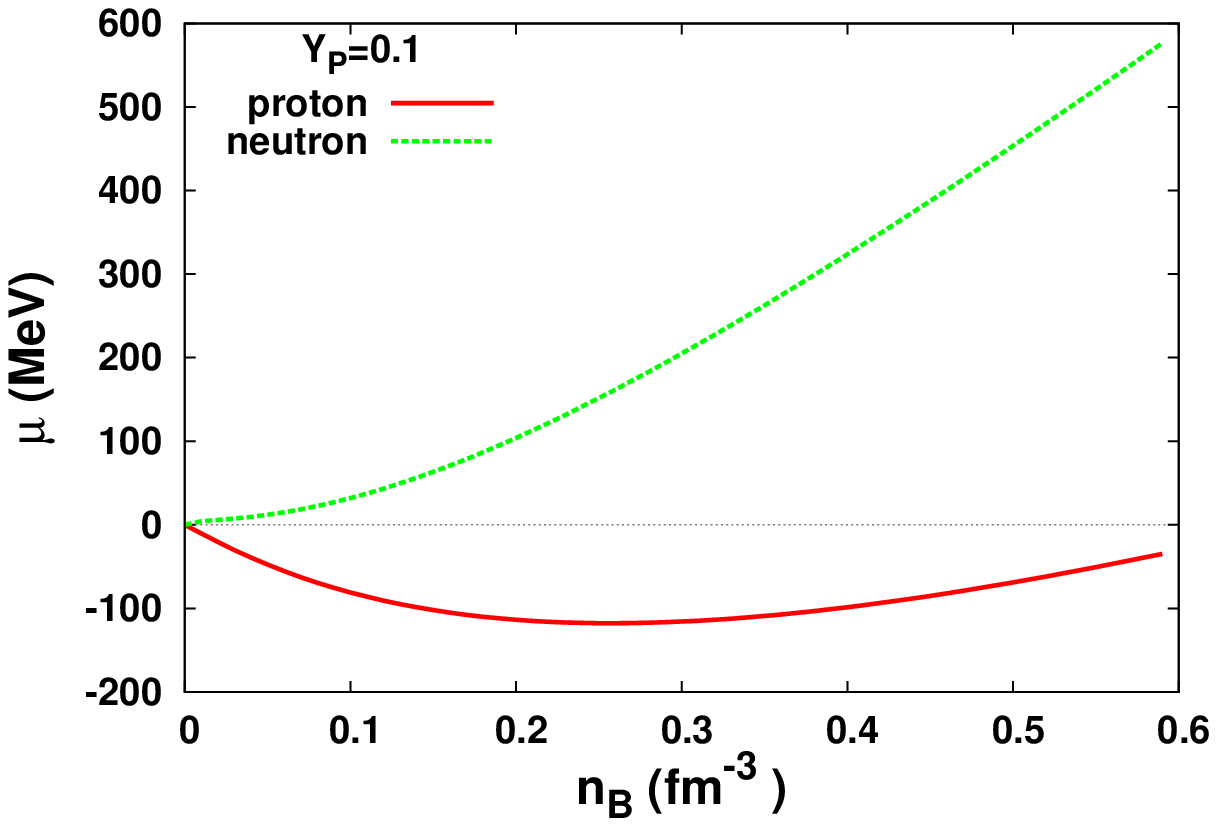}\\
\includegraphics[width=8.cm,angle=0]{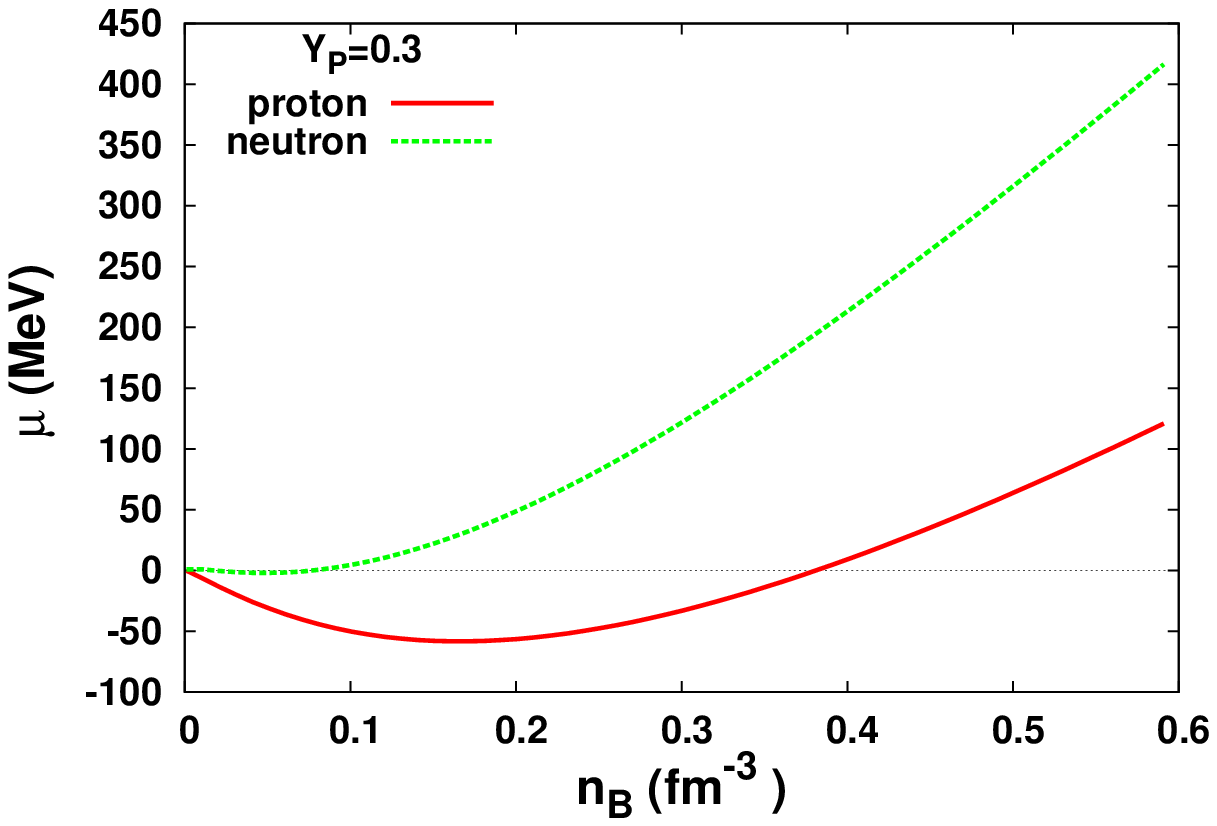}\\
\includegraphics[width=8.cm,angle=0]{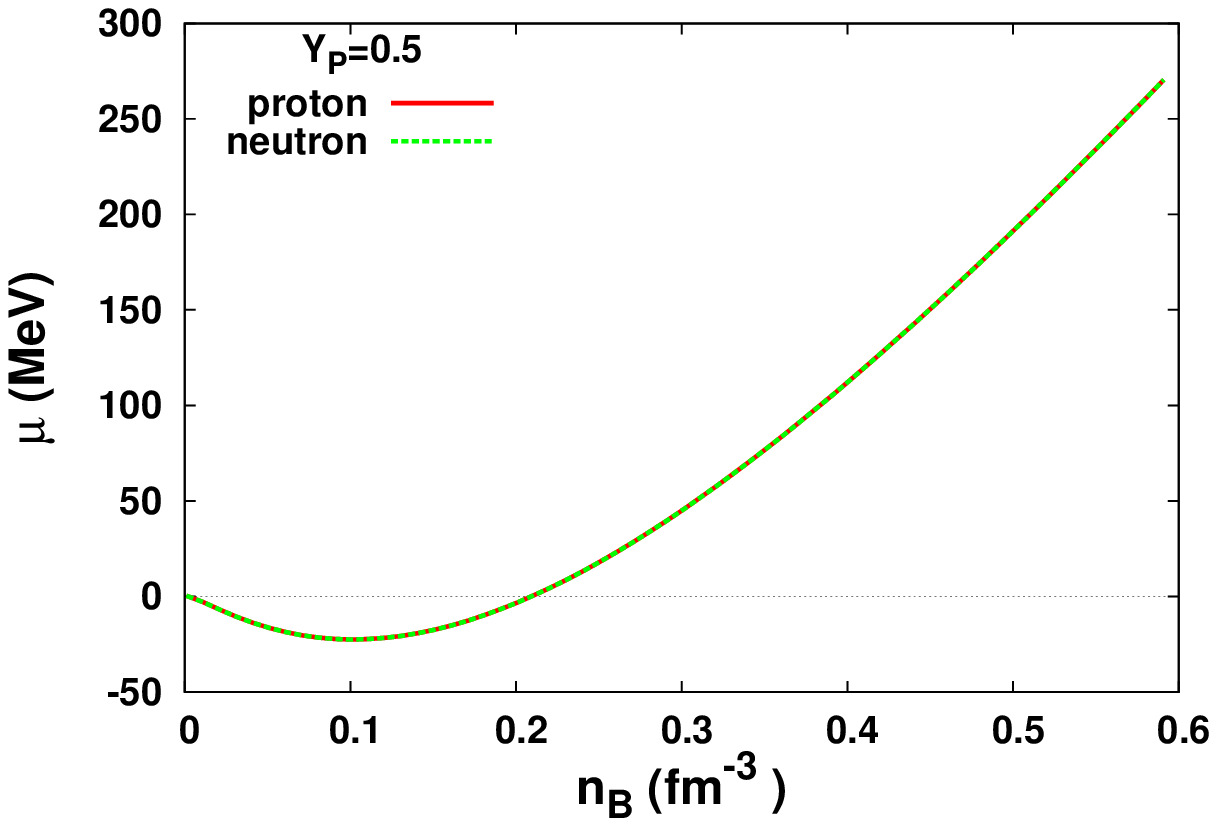}\\
\includegraphics[width=8.cm,angle=0]{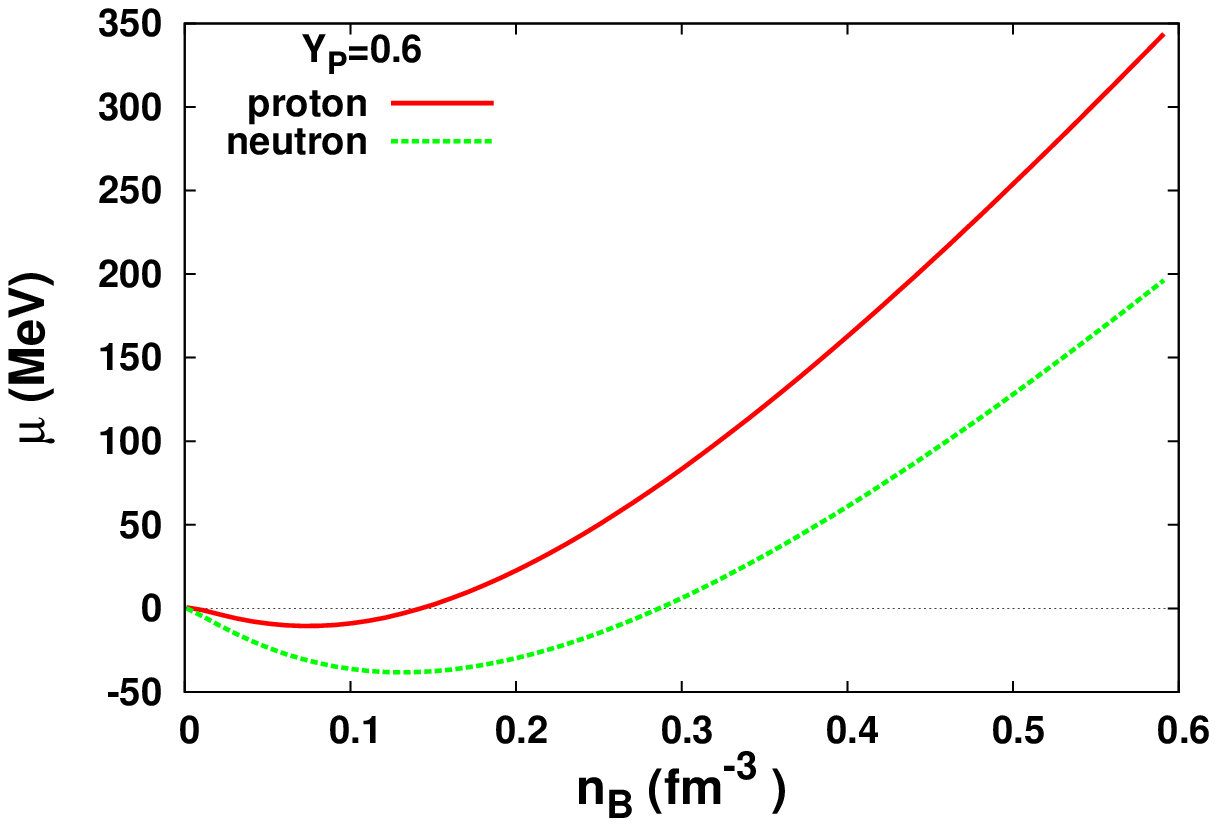}
\end{tabular}
\caption{Some fixed values of proton fraction for chemical potentials as function of the baryonic density. The continuous line represents $\mu _p$ and the dashed line represents $\mu _n$.}
\label{fig6}
\end{figure}


In Table \ref{tab:comp} we show the properties of nuclear matter at saturation density and zero
temperature. The first column shows the binding energy per nucleon, the second shows the 
baryonic density of the nucleons at the saturation point, the third the effective mass relative to the free nucleon mass, 
the symmetry energy and the compression modulus of the nucleons. 

In Fig. \ref{fig1} we show the relation of the effective mass as function of the baryonic density
where we can see that the effective mass for the QMC model has always a larger value than the 
other models, which is an important characteristic if hyperons are to be included. As can be seen in \cite{alex04}
the effective mass of some of the Walecka-type models tend to zero at still low densities when all the baryons of the octet are included.
For this reason, parameterizations with larger effective masses were
proposed by Glendenning \cite{gledenning} (GM1, GL, etc) with the
specific purpose of applications to stellar matter.
Note that in both, figure and table, we can see that $ M^*/M=0.77 $ at the saturation density, which agrees with the results
derived from the non-relativistic analysis
of scattering of neutrons from lead nuclei \cite{johnson}, and that
this result is larger in comparison with the effective mass from many of the 
Walecka-type models. 

In Fig. \ref{fig2} we show the pressure versus energy density relation for infinite nuclear matter at $ T=0 $, 
where we see that the curve for QMC is similar to the one for TW, and softer than the GM1 and NL3. As mentioned in the Introduction, this can be
a deficiency for neutron star studies, but those effects on supernovae are still not analyzed.

\section{Results and discussions}

In this work we construct an EoS table covering a wide range of proton
fraction $Y_p $ and baryon density $ n_B $. We show in the paper some results
concerning the properties of matter with the QMC model and indicate the
website from where the full table can be download.

We show in Fig. \ref{fig3} the binding energy of homogeneous nuclear
matter at zero temperature as a function of the
 baryon density for different proton fractions.
For pure neutron matter and low proton fractions, there are no binding states, as expected. This result is 
in good agreement with Walecka \cite{walecka} and Shen \cite{shen98}.  
In Fig. \ref{fig4} we show the compression modulus versus the baryonic density, where the expression for $K$ is
\begin{equation}
K=9\; n_B^2 \frac{d^2}{dn_B^2}\left ( \frac{\varepsilon}{n_B} \right ),
\end{equation}
$n_B$ is the baryon number density and $\varepsilon$ the total energy density of the nucleons. Here we see that near the saturation density the compression modulus 
has the same value independent of the proton fraction and, as the density increases, the compression modulus is different for each $Y_p$. 

In Fig. \ref{fig5} the pressure $p$ as a function of $\rho_B$ is
displayed. The baryon number density is related to the baryon mass density as
$\rho _B=m_u n_B$ where $m_u =931.49432$ MeV is the atomic mass unit. We note that the pressure varies more with the $Y_p$ for 
low values of $p$ at lower densities, 
this result is also in good agreement with Shen's work \cite{shen98}. 	

In Fig. \ref{fig6} we show the proton and neutron chemical
potentials, $\mu _p$ and $\mu _n$, as function
of the baryon density for the proton fractions $Y_p=0.1$, $0.3$, $0.5$ and $0.6$. In these curves we see that for $Y_p=0.1$ the chemical potential of the neutron
is bigger than the one of the proton. As the proton fraction gets bigger, the curves approach each other, until they are the same in $Y_p=0.5$, and for $Y_p=0.6$
the chemical potential of the proton is bigger than the one of the
neutron. This is an obvious result, but as the chemical potentials are
very  important quantities in the EoS tables, they are also presented graphically.

\begin{table*}[t]
  \centering
	
  \begin{tabular}{lclcccccccccccc}
\hline
  T   & log$_{10}$$(\rho_B)$& $n_B$      & Y$_P$  &   F    &$E_{int}$& S     & M*$_N$   & X$_n$  & X$_p$ &     p         & $\mu _n$ & $\mu _p$ \\
 (MeV)&    (g.cm$^{-3}$)    & (fm$^{-3}$)  &      & (MeV)  & (MeV)   & (k$_B$) &  (MeV) &      &     & (MeV fm$^{-3}$) &  (MeV) & (MeV)  \\
\hline
 0    & 14.0              & 0.0602     &  0   & 4.890  & 12.40   &   0   &  838.9 &  1   &  0  &   0.2371      & 23.43  & -68.43  \\
 0    & 14.1              & 0.0758     &  0   & 6.090  & 13.60   &   0   &  816.9 &  1   &  0  &   0.5103      & 31.20  & -82.21 \\
 0    & 14.2              & 0.0954     &  0   & 8.133  & 15.64   &   0   &  791.2 &  1   &  0  &   1.0890      & 42.68  & -97.63 \\
 0    & 14.3              & 0.1201     &  0   & 11.56  & 19.07   &   0   &  761.5 &  1   &  0  &   2.2780      & 59.66  & -114.3 \\
 0    & 14.4              & 0.1512     &  0   & 17.19  & 24.70   &   0   &  728.8 &  1   &  0  &   4.6550      & 84.64  & -131.4 \\

\hline
  \end{tabular}
\caption{EoS table at $ T=0 $. It covers the proton fraction range $Y_p=0 - 0.65$ with the linear grid spacing $ \Delta Y_p=0.01 $ (66 points), and the density
range $ \rho_B=10^{14} -10^{16}$ g cm$^{-3}$ with the logarithmic grid spacing $\Delta $log$_{10}$ $(\rho_B$ /[g cm$^{-3}])=0.1$ (21 points).
This table is available in the website . An exerpt of it is shown here for guidance.}
  \label{tab:table}
\end{table*}

Once bulk nuclear matter properties are shown to behave as expected and present some important differences as compared with the other works,
we proceed toward building a preliminary EoS table with the QMC model, for homogeneous matter and zero temperature, which is available on the Web at
\\

\url{http://debora.fsc.ufsc.br/eos_qmc.t0}.
\\

In Table \ref{tab:table} we show the thermodynamic quantities described as in \cite{shen98,shen11}

\begin{enumerate}

\item Temperature: $T$[MeV].

\item Logarithm of baryon mass density: log$_{10}$($\rho _B$)[g.cm$^{-3}$].

\item Baryon number density: $n_B$ [fm$^{-3}$].\\

\item Proton fraction: Y$_p$.\\
The proton fraction Y$_p$ of uniform matter made of protons and neutrons, is defined by
$$
Y_p=\frac{n_p}{n_n + n_p}
$$
where $n_p$ and $n_n$ are the number density of protons and neutrons, respectively.

\item Free energy per baryon: $F$ [MeV].\\
The free energy per baryon reads,
\[
f=\varepsilon-Ts.
\]
This work is for zero temperature only, hence $f=\varepsilon$. The free energy per baryon is defined
relative to the nucleon mass as,
\[
F=\frac{\varepsilon}{n_B}-M=B/A.
\]

\item Internal energy per baryon:$E_{int}$ [MeV].\\
The internal energy per baryon is defined relative to the atomic mass unit $m_u =931.49432$ MeV as
\[
E_{int}=\frac{\varepsilon}{n_B}-m_u.
\]
\item Entropy per baryon: $S$[$k_B$].\\
The case we work here is for zero temperature, therefore $S=0$.

\item Effective nucleon mass: $M^{*}_N$ [MeV].\\
The effective nucleon mass is obtained in the QMC model for uniform matter with the relation
$M^{*}_N=E^{bag}_N$, where $N=p$, $n$, and the bag energy is obtained through the Eq. \ref{ebag}.

\item Free neutron fraction: $X_n$.

\item Free proton fraction: $X_p$.

\item Pressure: $p$ [MeV.fm$^{-3}$].\\
The pressure is calculated from Eq. \ref{press}.

\item Chemical potential of the neutron: $\mu _n$ [MeV].\\
For the zero temperature case, the chemical potential of the neutron relative to the free nucleon mass $M$
is calculate through the following equation
\[
\mu_n=[k_n^2 +M^{*2}]^{1/2}+g_\omega \omega_0 -\frac{g_\rho}{2}\rho_{03}-M.
\]

\item Chemical potential of the proton: $\mu _p$ [MeV].\\
For the zero temperature case, the chemical potential of the proton relative to the free nucleon mass $M$
is calculated through the following equation
\[
\mu_p=[k_p^2 +M^{*2}]^{1/2}+g_\omega \omega_0 +\frac{g_\rho}{2}\rho_{03}-M.
\]

\end{enumerate}

\section{Conclusion and future works}

In this work we have used the QMC model for the first time for the 
construction of a preliminary EoS that in the future can be useful for
the studies involving  cooling of neutron stars and supernova simulations.
We believe that with the quarks degree of freedom present in the QMC
model, the EoS can contribute with part of the
physics that lack for SN simulations to explode.

The next step of the present  work, already under development, is the
computation of the EoS grid at finite
temperature, which is essential for the supernova simulations. 

Thereafter we will study the very low
density regions, where nuclear matter is no longer uniform. This will
be done with the {\it pasta phase} approach \cite{maruyama,constanca}. We believe that the use
of the pasta phase for the description of the non-uniform part of matter that compose
the EoS table  in the SN simulations will certainly affect SN and
cooling simulations.

%
\section*{ACKNOWLEDGMENTS}
The authors would like to thank CNPq and FAPESC under the project 2716/2012, TR 2012000344 for the financial support.
\bibliography{biblio.bib}
\end{document}